\documentclass[reprint,
 amsmath,amssymb,
pra,
]{revtex4-2}
\usepackage{url,hyperref,lineno,microtype,subcaption,listings,xcolor,graphicx}
\usepackage{booktabs}

\definecolor{codegreen}{rgb}{0,0.6,0}
\definecolor{codegray}{rgb}{0.5,0.5,0.5}
\definecolor{codepurple}{rgb}{0.58,0,0.82}
\definecolor{backcolour}{rgb}{0.95,0.95,0.95}

\lstdefinestyle{mystyle}{
    backgroundcolor=\color{backcolour},   
    commentstyle=\color{codegreen},
    keywordstyle=\color{magenta},
    numberstyle=\tiny\color{codegray},
    stringstyle=\color{codepurple},
    basicstyle=\ttfamily,
    breakatwhitespace=false,         
    breaklines=true,                 
    captionpos=b,                    
    keepspaces=true,                 
    numbers=left,                    
    numbersep=5pt,                  
    showspaces=false,                
    showstringspaces=false,
    showtabs=false,                  
    tabsize=2
}

\begin{document}
%\firstpage{1}

\title[Scaling Scientometrics]{Scaling Scientometrics: Dimensions on Google BigQuery as an infrastructure for large-scale analysis} 

\author{Daniel W Hook}
 \altaffiliation[Also at ]{Centre for Complexity Research, Imperial College London, London, SW7 2AZ, UK and Department of Physics, Washington University in St Louis, St Louis, Missouri, US.}
\author{Simon J Porter}%
 \email{s.porter@digital-science.com}
\affiliation{%
 Digital Science, 6 Briset Street, London, EC1M 5NR
}%

\begin{abstract}
Cloud computing has the capacity to transform many parts of the research ecosystem, from particular research areas to overall strategic decision making and policy.  Scientometrics sits at the boundary between research and the decision making and evaluation processes of research.  One of the biggest challenges in research policy and strategy is having access to data that allows iterative analysis to inform decisions.  Many of these decisions are based on ``global'' measures such as benchmark metrics that are hard to source.  In this article, Cloud technologies are explored in this context.  A novel visualisation technique is presented and used as a means to explore the potential for scaling scientometrics by democratising both access to data and compute capacity using the Cloud.
\end{abstract}

\maketitle

\section{Introduction}
In recent years cloud technologies have become used more extensively in research.  The combination of cost-efficient storage and on-demand compute capability have lowered barriers for many who are either not technically savvy or who lack financial resource to create and maintain large scale real-world computer infrastructure.  In the academic discplines of bibliometrics and scientometrics, and in the related practical fields of research management, strategy and policy, the use of cloud-based tools are still naiscent.  On one hand data volumes are relatively small (at least compared with familiar big data fields such as particle physics) while on the other, the costs and complexity of arranging access to bibliometric data sources, processing raw data and maintaining analysis-ready datasets have been prohibitive for all but the best funded researchers, analysts and policymakers.

We argue that Cloud technologies applied in the context of scientometrics does not only have the capacity to democratise access to data but also to democratise access to analysis.  Here we define ``analysis'' to be the combination of data access together with the capacity to calculate.  Data access is often thought to be constrained solely by licence agreements, but is also characterised by technical limitations.  Recent progress has been made in improving access to research metadata \citep{ludo_waltman_open_2020}. Yet, data licence agreements typically do not make arrangements for the delivery of an often-updated analysis-ready database, but rather give access either to a raw flat-file data that needs to be processed, structured and mounted into a database format with regular updates that must be applied to keep the data relevant, or access to an API, which must go through a similar process to create an analysis-ready database.  Beyond this logical data structuring activity, there has also historically been the need for physical hardware, that effectively defines the computational capacity of the user.  Cloud technologies have the capacity to remove both of these constraints by providing an analysis-ready database and computational capacity on a per-use basis.

Few research areas yet take an approach of providing a Cloud-based central store of research data for researchers to query, manipulate and compute with to support their investigations.  However, this type of approach can be seen in the conception of ``computable data'' introduced by \cite{wolfram_making_2010} as a result of the development of Wolfram Alpha.

In this article we seek to highlight the types of analysis that can be carried out if data is made accessible in the Cloud, as described above, as well as the implications for community-ownership of research benchmarks, and the opportunity to place analytical capabilities with a far broader range of stakeholders. 

To begin, we provide a working definition of accessibility and use Dimensions on Google Big Query to explore a simple example related to the field of ``knowledge cartography'', which was introduced and explored extensively by \cite{borner_visualizing_2003,boyack_mapping_2005,boyack_mapping_2007,borner_atlas_2010,borner_design_2012,borner_atlas_2015}.  We use this example as it has great narrative power and makes global use of a dataset.  (Here, by global, we mean that to complete an analysis every record in the dataset must contribute toward the result---a good example of a global calculation is a field-weighted citation normalisation, since this requires the citation counts of every publication in a set for a defined time period.)

This example brings together the use of a structured, analysis-ready dataset hosted on the Cloud, with unique identifiers to connect from metadata records to spatial information with on-demand computation to provide a visualisation that can readily be updated, iterated and provided regularly to stakeholders in a maintainable manner.  We believe that the analysis presented here is entirely novel in a bibliometric or scientometric context. It is remarkable that results of this type have not been presented by other researchers, but we take this to be a hallmark of the limitations of prior computational approaches.

\subsection{Defining data accessibility}
The viability and trustworthiness for bibliometric datasources has been a matter of significant attention in the bibilometrics community over recent years \citep{lopez-illescas_comparing_2009,garciaperez_accuracy_2010,mongeon_journal_2016,bornmann_field_2018,herzog_response_2018,martin-martin_google_2018,van_eck_accuracy_2019,huang_comparison_2020}.  The emergence of new datasources has led to significant analytical efforts to understand the strengths and weaknesses of different approaches to collecting and indexing content \citep{powell_coverage_2017,thelwall_dimensions_2018,martin-martin_google_2020,visser_large-scale_2020}.  The primary focuses of these works are in the assessment of coverage (completeness of journal/subject coverage, and accuracy and completeness of citation network) together with technical issues around stable construction of field normalisations and other benchmarking details.  Both of these areas are foundational in whether a database can be used in bibliometric and scientometric anaylsis, and whether it is appropriate to use these data in evaluative contexts.  More recently, there has been innovative work which extends this standard approach to assess coverage in a different manner to examine suitability of datasets for ``bibliometric archeology'' \cite{bornmann_growth_2020}.

For the purposes of this paper, we characterise the majority of existing comparative analyses as being focusing on one or more of five key data facets: 
\begin{enumerate}
 \item coverage--the extent to which a body of metadata covers the class of objects that it sets out to catalogue; explanations of editorial decisions, limitations based on geography, subject or nature;
 \item structure--the format and field structure of the metadata; the standards which may be relevant;
 \item nature--the parts of the scholarly record being captured (e.g. articles, journals, datasets, preprints, grants, peer reviews, seminars, conference proceedings, conference appearances, altmetric information, and so on); level of granularity; 
 \item context--provenance; details of enhancement techniques that may have been applied; use of AI or machine-learning algorithms used;
 \item quality--data homogeneity; field completeness; completeness of coverage; quality of sourcing -- how easily can a calculation be performed and how reliable is the resulting analysis?
\end{enumerate} 

The first four of these aspects of a dataset define the extent of a ``data world'' that may be explored and analysed to deliver insight. If we wish to push out the boundaries of this world, then we can do that by improving each of these facets: Extending the coverage of the database, deepening sophistication of the facetting, expand the different types of data that we include for analysis, or by broadening the links between different parts of the data to improve context.  Data quality determines the accuracy of how view of this landscape and the level of trust that we can have in analyses.

It may be arged that more established data sources have sought to optimise coverage, structure and quality of their data.  But, newer databases have brought a new focus on nature and context \citep{hook_dimensions_2018,herzog_dimensions_2020}. By expanding the types of data they that cover, or by creating better linkages between those new data types to improve our ability to contextualise data, they improve the variety and subtlty of the insights that the scientometrics community may generate.  We do not suggest that our list of analytical facets that drive value is an exhaustive one.  There are many additional features that change the value of any analysis such as considerations outside the dataset such as the affiliation of the analyst, or technical considerations such as data homogeneity, or robustness of statistical treatment.  We argue that data accessibility is a different type of feature of a dataset that should be considered more actively, especially in the rise of cloud technologies.

Data accessibility is a complex and multifaceted topic.  The key facets that we believe to be important in the context of scientometric data and analysis are:
\begin{enumerate}
    \item Timeliness: the extent to which a user can access sufficiently up-to-date data;
    \item Scale: the extent to which it is possible to access the whole dataset for calculational purposes;
    \item Equality: the extent to which the resources to process the data and perform a calculation are technologically practical;
    \item Licence: the legal terms that define the extent to which the data may be used and published.
\end{enumerate}

The example that we use here does not attempt to illustrate or address all these facets.  Recent work by \cite{hook_real-time_2020} focused on timeliness.  In the current article we focus on scale and equality. Specifically,  we examine classes of calculation for which data access is required for scale and look at how Cloud technologies can facilitate both scale and equality of access to data. Our example will use Digital Science's Dimensions on BigQuery infrastructure. We note that this paper is specifically designed not to be a comparative study of the accessibility of different data sources, but rather as an opportunity to showcase the types of analysis that can be carried out if technological choices are made that democratise data access.

This paper is organised as follows: In Sec.~\ref{Sec:Method} we describe the Dimensions on Google BigQuery technical stack, and the specific queries used for the analysis presented in the following section.  In Sec.~\ref{Sec:Results} we show the results of several different calculations of the centre of gravity of research production using the method described in Sec.~\ref{Sec:Method} and discuss the context of those results.  In Sec.~\ref{Sec:Discussion}, we consider the potential of Cloud technologies to meet a broad set of use cases.

\section{Method}
\label{Sec:Method}
\subsection{Technical Infrastructure}
\label{Sec:Tech}
Many Cloud technologies are already used across research, especially in technical subjects requiring large-scale computation or storage, or those who engage in large scale collaborations.  Indeed, Cloud technologies are becoming more widespread in research as they prove to be highly cost-effective for some types of research activity. Typical use cases involve storage and transfer of data or obtaining computational power on demand.

For those with structured data, the Cloud technologies that allow users to not only store and distribute access to a dataset but also to perform complex calculations with an on-demand infrastructure are now coming of age. Technologies such as Amazon Redshift, Snowflake and Google BigQuery all have the potential to meet the use cases mentioned above \citep{zukowski_cloud-based_2018}.  

In addition to their technical capabilities, these technologies are opening up new business models through the ability to share secure data in a fine-grained and controlled manner.  Any of the technologies mentioned allows a data holder to share data from their Cloud database with others on a permissioned basis, opening up access specifically or generally based on many different criteria.  From a business model perspective, a critical differentiator (not used in the current example), is that two parties can add their data to the cloud completely securely, one can keep their data private while the other can open their data up on some mix of open access and commercial basis.  The second actors data can then be used by the first actor, on whatever the appropriate contractual terms are, mixing the data with their private data in a completely secure manner. The only requirement is that each dataset should have a sufficient overlap in persistent unique identifiers to allow the datasets to be compatible.  Hence, this technology is a strong reason for all stakeholders in the community to adopt and ensure that the data that they expose is well decorated with open identifiers. For large, frequently updated datasets where there is significant overhead in just storing and updating the data, this new way of working compeletely changes the basis of engagement.

From the perspective of the current article, the availabilty of Dimensions data in the Google BigQuery Cloud environment allows users to access and compute directly with the data without having to invest in either building or maintaining local infrastructure.  It also allows users to manipulate and calculate with the data across the whole Dimensions dataset.  The only technical expertise that is required is an ability to program with SQL.

It is easy to see how the calculation explained below could easily be replaced to calculate other metrics and indicators that require access to a ``global'' dataset.  Such calculations include journal metrics such as Journal Impact Factor \citep{garfield_new_1963}, EigenFactor \citep{bergstrom_eigenfactor_2007}, SJR \citep{gonzalez-pereira_new_2010} or CiteScore \citep{van_noorden_controversial_2016}, as well as the production of journal citation distributions \citep{lariviere_simple_2016}, field-based normalisations such as RCR \citep{hutchins_relative_2016}, as well as geographical benchmarks, trend analysis or examples of knowledge cartography, such as the example that we have chosen to explore.

\subsection{Calculation for Example}
\label{Sec:Calc}
To illustrate how the new technologies described above may be used, we perform a simple global calculation.  As noted above, the word ``global'' here is not intended to refer to a geographical context, but rather implies that each record in the database will potentially contribute to the calculation.  

We calculate the centre of mass of global research output year by year.  This calculation has several noteworthy features that demonstrate the capabilities that we've discussed earlier.  The calculation: i) involves every publication record in our dataset; ii) makes use of a unique identifier to connect publication outputs to geographical locations (in our case through GRID); iii) makes use of the time-depth of the publications records in the database to give a trend analysis.

Using non-Cloud infrastructure to perform this calculation such as a standard relational database hosted on physical infrastructure would make this calculation time consuming and resource intensive.  By leveraging Cloud infrastructure we can quickly iterate the detail of this calculation to test different hypotheses.  For example, we can easily shift from a centre of mass calculation that focuses on publications to one that focuses on awarded grants, patents or policy documents.  We can trivially change the weighting factor from an unweighted calculation to a citation weight in the case of publications, grant size in USD for grants, the funded associated with a publication, the altmetric attention associated with a patent and so on.  We can also easily restrict our analysis to a specific research topic, country, institution, a specific class of grants, a particular type of funding or a larger-scale policy initiative such as open access.  To take this even further, one can imagine even subtler weighting schemes that take the CRediT taxonomy \citep{allen_publishing_2014} into account.

In the examples contained in this paper we focus on publication output and either unweighted or citation-weighted formaulations.  The core of the centre of mass calculation is a simple weighted average of spatial positions that all students of classical mechanics meet early in their studies - it is equivalently known as a centre of gravity calculation or centroid.  

In our example, each ``mass'' is an affiliated research institution and the location of that mass is the geographical location of the principle campus as recorded in GRID.  For each individual paper, there is a centre of mass the position of which is proportional to the contribution of the affiliations of the researchers who have contributed to the paper.  For exmaple, if a paper were to be entirely written from researchers at a single institution then the centre of mass for the paper in our calculation would be the location of the principle campus in GRID.  If a paper were to be written by two co-authors, one at the University of Cambridge and the other at the University of Oxford, then the centre of mass would be computed to be midway between the two Senate House buildings of the two institutions.  To find the centre of mass of global output in any year, we average the spatial location of all the papers produced in that year. We can think of this position as the ``average centre of global research production'' or the ``centre of mass/gravity of global research output''.  

We also introduce a citation-weighted version of this calculation which may be interpreted as a measure of centrality of global research attention to research output.

Formally, we define the centre of mass of a set of research objects to be the spatial average (or centroid) of the affiliations of the co-creators of the output.  On a paper with $n$ co-authors, each co-author is associated with $1/n$ of the paper.  If a given co-author is affiliated with $m$ institutions, then each institution will have a weight of $1/m$ of that co-author's part of the paper, and $1/nm$ of the overall paper.  Thus, each author-institution pairing has a weight $a_{nm}$ where 
\begin{equation}
    \sum_n\sum_m a_{nm}=1.
\end{equation}
We do not need to explicitly sum over authors to get the overall contribution of a specific institutions nor do we need to worry about repetition of institutions since, in our calcualtion, we reduce an institution to the longitute and latitude of its principal campus.  Hence, there is a natural accumulation of weight to a geographical location.

This reduction to longitude and latitude is made possible through the use of GRID.  The longitude and latitude of research institutions is not held natively within the Dimensions dataset.  However, each institution in Dimensions is associated with a persistent unique identifier that allows us to connect to other resources.  In the case of Dimensions the institution identifier is the GRID identifier.  GRID not only includes some helpful data about institutions such as the longitude and latitude that we use here but also acts as a gateway to resources such as ROR (the Research Organisation Registry) that will in turn facilitate access to other pieces of information.

This means that we can simply calculate the average longitude, $\overline{long}$ and latitude $\overline{lat}$ of a single research output using:
\begin{eqnarray}
  \overline{lat} &=& \frac{1}{T}\sum_i \sum_j lat_{ij};\nonumber \\ \overline{long} &=& \frac{1}{T}\sum_i \sum_j long_{ij},
\end{eqnarray}
where $T$ is the total number of publications.

We can then extend this to a group of outputs by introducing an index, $k$, that ranges over each output in the relevant set to create the average longitude $\overline{Long}$ and average latitude $\overline{Lat}$ of the whole set:
\begin{eqnarray}
  \overline{Lat} &=& \sum_k\frac{1}{T_k}\sum_i \sum_j lat^k_{ij}; \nonumber \\ \overline{Long} &=& \sum_k\frac{1}{T_k}\sum_i \sum_j long^k_{ij},\label{eq3}
\end{eqnarray}
where $T_k$ is the total number of institutional affiliations on the $k^{th}$ paper in the average.

Longitude and latitude are defined as angles on the surface of a sphere with longitude in the range $[-90,90]$ and latitude in the range $[-180,180]$. The construstion in Eqn.~\ref{eq3} guarantees that the final results of these calculations are also in these ranges.

Further weighting factors can also be added to the calculation to highlight issues of particular interest.  For example, if we were to consider an example using research publications and we wished to calculate not just the centre of the output rate but rather the centre of the combination of output weighted by the attention given to that output, then we might introduce a weighting by the number of citations received by each paper.  

In that case (\ref{eq3}) would need to be updated and the form for the centroid would be:
\begin{eqnarray}
  \overline{Lat} &=& \frac{1}{C}\sum_k\frac{C_k}{T_k}\sum_i \sum_j lat^k_{ij}; \nonumber \\ \overline{Long} &=& \frac{1}{C}\sum_k\frac{C_k}{T_k}\sum_i \sum_j long^k_{ij},\label{eq4}
\end{eqnarray}
where $C_k$ is the number of citations of $k^{th}$ paper and $C$ is the sum of all citations across papers in the set.

Likewise, if we were interested in the level of non-scholarly attention we might replace citations by some relevant altmetric data.

The code snippet below is the implementation of Eqn.~\ref{eq4} using Google BigQuery's implementation of SQL on the Dimensions dataset. In addition to the calculation explained above, the code below takes into account cases where creators may miss an affiliation by ensuring that the normalisation is consistent in the case of null data.

\begin{lstlisting}[language=SQL,caption={\label{lst:1}Listing to produce a citation-weighted centre of mass year-by-year using SQL on Google BigQuery with Dimensions data.},captionpos=b]
WITH pubs_reweighted AS (SELECT p.id,
         p.year,
         a.first_name,
         a.last_name,
         a.initials,
         /* count the distinct number of organisations per author */
         COUNT(distinct g.id) num_orgs,
         /* list of all the GRIDs per author */
         ARRAY_AGG(grid_id) grids,
         /* count the number of authors on the paper that have    affiliations in GRID */
         COUNT(p.id) over(partition by p.id) authors
FROM
     `dimensions-ai.data_analytics.publications` p
      INNER JOIN unnest(authors) a
      INNER JOIN unnest(a.affiliations_address) aff
      INNER JOIN `dimensions-ai.data_analytics.grid` g
         ON g.id = aff.grid_id
GROUP BY
        p.id,
        p.year,
        g.name,
        a.first_name,
        a.last_name,
        a.initials
),
/* get the location for each GRID. Each row that is being summed here represents a single author.  If they have more than one affiliation then the contribution of the author is split equally. */
pub_center_mass AS
    (SELECT pr.id,
          SUM((g.address.latitude/pr.num_orgs)/pr.authors)  latitude,
          SUM((g.address.longitude/pr.num_orgs)/pr.authors)  longitude
    FROM pubs_reweighted pr,
        UNNEST(grids) grid_id
        INNER JOIN `dimensions-ai.data_analytics.grid` g
             on g.id = grid_id
    GROUP BY pr.id)
SELECT p.year,
/* sum the centre mass for all publications / the number of publications; replacing p.metrics.times_cited in the respective to an explicit value of "1" recovers a weighting-free from the calculation */
      (sum(cm1.latitude * p.metrics.times_cited)) /   (sum(p.metrics.times_cited)) latitude,
      (sum(cm1.longitude *  p.metrics.times_cited)) / (sum(p.metrics.times_cited)) longitude
  FROM pub_center_mass cm1
   INNER JOIN `dimensions-ai.data_analytics.publications` p
      ON p.id = cm1.id
  GROUP BY p.year
  HAVING sum(p.metrics.times_cited) > 0
  ORDER BY year
\end{lstlisting}

One assumption that may not at first appear obvious with the weighted approaches used here is that the sum of all citations in time has been used.  As a result, papers in 1671 have had 350 years to garner citations whereas more recent publications have had much less time.  Of course, the average in each case is performed on a homogeneous basis (i.e. only publications of the same year are averaged together), however, this does introduce an implicit bias in the analysis in that a citation bias may have a comtemporary skew.  A further analysis could be performed that only considered the citations in an $n$-year window following the date of publication of the paper.  Of course, introducing such a parameter also makes a value judgement about the lifetime of a piece of research.

In Sec.~\ref{Sec:Results} we use this method to showcase three analyses: i) a standard unweighted calculation of the centre of mass of research output from 1671 to present day; ii) a calculation of the centre of mass of research weighted by citation attention over the same time period; iii) a calculation of the citation-weighted centre of mass of research based just on data from the freely available COVID-19 dataset that is available on the Google BigQuery environment.

\subsection{Data specifics}
The details of the high-level data schema in Dimensions, including information about coverage and the treatment of unique identifiers is described in several recent publications, for example \cite{hook_dimensions_2018,hook_real-time_2020}.

Once the data are produced from a script such as the one above they were downloaded from the interface and are initially analysed in Mathematica.  The graphics shown in Sec.~\ref{Sec:Results} are produced using Ddtawrapper.de.

\begin{figure}[!h]%
\begin{center}
\includegraphics[width=0.8\linewidth]{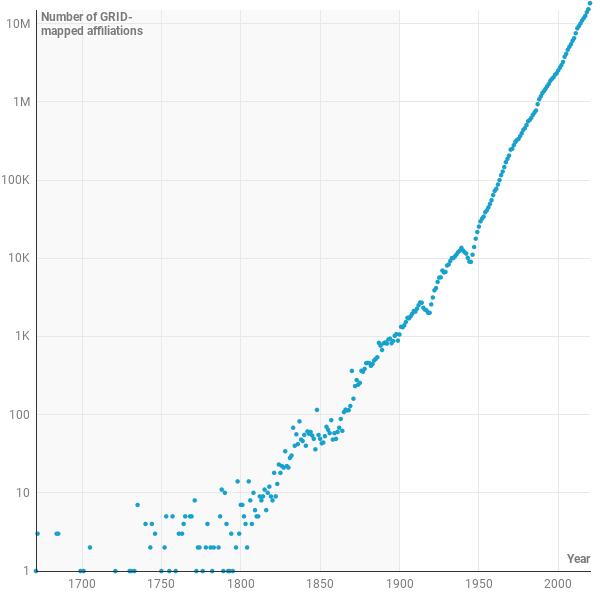}%
\end{center}
\caption{Logarithmic-scaled plot of the number of GRID-mapped institutions associated with papers in the Dimensions database by year from 1671 to 2020.  The two notable dips in the data in the first half of the 20th Century co-incide with the two world wars.  The grey background highlights the region between 1671 and 1990 in which the number of contributing records is taken to be too small to give a stable basis for statistical analysis.}%
\label{Fig:1}%
\end{figure}

At the Mathematica analysis stage, we plotted every year of data from the system.  However, this gave an unsatisfactory picture as the data are quite messy.  In the early years of the dataset (approximately from 1671-1850) the number of publications with a GRID-listed institutions number in the single digits.  A confluence of reasons contribute to this picture: i) the low number of overall publications; ii) the low level of stated academic affiliations of authors in early work; iii) affiliations to institutions that are not part of GRID.  Figure~\ref{Fig:1} shows the number of publications with at least one recognisable (GRID-mapped) affiliation in each year in the Dimensions dataset.  

From 1900, the data begins to settle as it begins to be appropriate to treat it statistically in the context of a statistical calculation such as the one outlined in Sec.~\ref{Sec:Calc}. Between 1900 and 1970, the year-on-year variability of the data decreases, and from the 1970s the data describes a fairly consistent path with few significant derivations. As such, we have denoted points in the figures in grey where they contain ``less robust'' data and in red when the data are ``more robust''.

\begin{table}
\begin{center}
\begin{tabular}{@{}lr@{}}
\toprule
\multicolumn{1}{c}{\textbf{Month}} & \multicolumn{1}{c}{\textbf{Number of publications}} \\ \midrule
January & 289   \\
February  & 751 \\
March  & 3,140   \\
April  & 9,999   \\
May & 15,502     \\
June  & 15,377   \\
July  & 16,706   \\
August  & 15,645 \\
September  & 16,191  \\
October    & 18,304  \\
November & 15,170    \\
December  & 15,153   \\ \bottomrule
\end{tabular}
\end{center}
\caption{Number of COVID-19 research publications including journal articles, preprints, monographs and book chapters by month during 2020 in the Dimensions database.}
\label{tab:t1}
\end{table}

In the final analysis presented, we focus on the COVID-19 dataset and perform a month-by-month analysis.  In this situation, we are again in the law of relatively small numbers where we have to be careful about statistical effects. However, the COVID-19 dataset has grown quickly during 2020 with a few hundred papers in January growing to several thousand papers per month in November (see Table \ref{tab:t1}).

\section{Results}
\label{Sec:Results}
From a historical perspective, the calculation of a variety of difference centres of mass can be revealing.  At the least, they may confirm accepted doctrine, but in the best situation they can reveal features that allow us to quantify and understand how aspects of our society are developing in a very relatable manner. 

Bibliometric analyses such as those presented here have previously been difficult to undertake due to the challenges of arranging data access, having the capacity to process data into an appropriate format, having the computation capacity to perform calculations and having a good reason to do put effort into generating this kind of output.  With the arrival of cloud-based technologies the technical challenges are removed.  A mere 40 lines of code, with a runtime of significantly less than 1 minute, is required to produce the data that underlies the analysis presented here based on the Dimensions dataset.  

By comparison such plots are relatively more common in other areas of research, such as economics or geography.  The recent work of \cite{dobbs_urban_2012} examined the movement of the centre of mass economic activity in the world from 1CE to the present day, showing that the economic centre of mass 2 millennia ago lays on a line between Rome and China.  During this period, the Silk Roads was the commercial axis between the two largest empires in the world: the Roman Empire and the Eastern Han Empire.  It is unsurprising that the economic centre of gravity is closely linked to these ancient centres of commerce.  The centre of mass was solidly grounded in the same region until at least 1500.  However, following the englightenment in the 18th Century, science and technology began to transform the economies of Europe and for a century from 1820 to 1913 the centre of mass of the world's economy moved rapidly West and North as the Industrial revolution transformed first the UK and then the wider Western world. Interestingly, in the McKinsey analysis, despite America's increasing world status and riches during the 20th Century, the centre of economic mass never quite left the eurasian continent, reaching its zenith in 1950, just over Iceland,  before beginning its journey Eastward and, again, Southward as first Europe emerged from war, Japan developed economically during the 1980s and finally China reached economic preeminence as we entered the Asian Century \citep{rachman_easternisation_2017}.

Most in academia agree that formal research publication dates from 1665 with the first issue of the Philosophical Transaction of the Royal Society \citep{hurst_trailblazing350_2010}.  Hence, the data that we have around research activity only spans a few hundred years and does not share the time-depth available in the work of \cite{dobbs_urban_2012}.  As a result, from a data perspective, we miss much of the detail around the development of older societies such as those in Egypt and China.  Anaecdotally, it is particularly interesting that the Chinese did not develop a research community with the associated communication structure despite significant technologies through the Ming and Qing periods.  Indeed, many of the principles that led to the Enlightment in Europe had parallels in Qing China and there is even evidence in European writings that they were aware of enlightnment-style developments in China \citep{wood_story_2020}. Yet, this does not appear to have resulted in the emergence of formal research publication culture. Miodownik offers a material scientist's view in \cite{miodownik_stuff_2014} on the relative rate of development of Chinese science - it may be that the development and wide adoption of glass in preference to porcelain is the small change that shaped the development of history for several centuries. 

The scholarly communications community has associated today's digital infrastructure (such as persistent unique identifiers) with pre-digital-era publications and this gives us an ability to piece together a much fulller picture than would otherwise be the case. Nevertheless, Figure~\ref{Fig:1} makes it clear that data are not sufficient to be treated in a reasonable statistical manner until much more recently. For the purposes of our example, we have chosen to keep the more statistically questionable points on our plot for aesthetic reasons, but have coloured these points in later figures in grey to denote the intrinsic uncertainty and arbitrariness of the choice of the data point.  

\begin{figure}[!h]%
\begin{center}
\includegraphics[width=.9\linewidth]{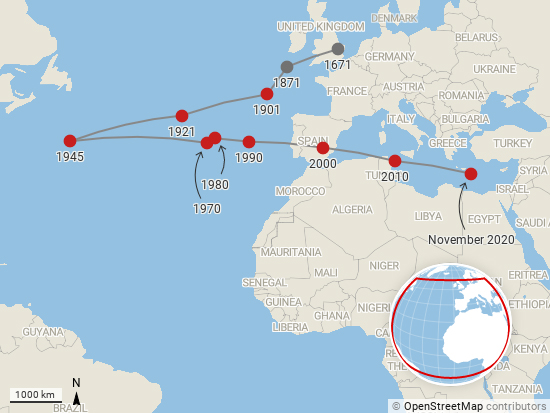}%
\end{center}
\caption{Motion of the centre of mass of research production from 1671 to present day.  The centre of mass calculation is unweighted by citations or other measures and is based solely on the outputs of papers by institutions that appear in the GRID database.}%
\label{Fig:2}%
\end{figure}

Figure~\ref{Fig:2} shows the motion of the unweighted centre of mass of global publication output between 1671 and the present day.  The start point of the path is an easy one to calculate since only one publication in that year is associated with a DOI and a GRID-resolved institution.  The paper concerned is a Letter that appeared in the Philosophical Transactions of the Royal Society of London.  It is written by ``Mr. Isaac Newton, Professor of the Mathematicks in the University of Cambridge; containing his new theory about light and colors''.  The path is highly volatile in the years following 1671 as the number of papers is small (those interested in this detail can review the annual calculation in the supplementary material). However, by 1901, there is are sufficiently many papers with well-identified institutions that the path settles somewhat.  

Many of the great academic institutions in the US had been established in the late 18th Century. Through the 19th Century the ``Robber Baron'' industrialists such as Mellon, Carnegie and Rochefeller had continued the trend of setting up academic institutions and by the 20th Century, these institutions were pulling the centre of mass of research (eratically at first, but then with increasing speed) away from Europe.  The First and Second World Wars saw significant disruption in Europe and the wealth that had taken the British Empire a century to accumulate travelled to the US in just four years as Britain underwrote the costs of the First World War between 1914-1918.  And so, the movement of the centre of mass of research production makes complete sense from 1900-1945.

If anything, it is remarkable that 1945, the year that Vannevar Bush wrote his famous Endless Frontier report \citep{bush_endless_1945}, marks the turning point of the transit of the centre of mass back toward Europe.  While the end of disruption in Europe meant that academics could return to their research and publication could begin again, Germany was in ruins and the economy of the UK was in tatters.  Despite the success of US-based programs such as the Manhattan Project during the war, research focus had yet to come to the fore in US universities.

Following Bush's report, the National Science Foundation was created and the formal basis for a period of US-centred scientific pre-eminence was established.  In Europe, the reorganisation of research was also under way, the Kaiser Wilhem Institute was renamed to the Max Planck Institute in 1948 and in 1949 the Frauhofer Institute was established.  By the 1960s, the Royal Society of Great Britain would coin the term ``Brain Drain'' to describe the movement of British Scientists from the Old World to the New \citep{cervantes_brain_2002,balmer_royal_2009}. In the UK, Wilson's White Heat of Technology of the 1960s \citep{morris_science_1993} served to help to keep the centre of mass moving torward Europe.  

Overall the balance of publication volume remained in Europe's favour from 1945 until 1970, with a slow draft in the centre of mass of publication toward Europe.  During the final decade of this period US spending on research as a proportion of its discretionary budget reached an all-time high \citep{white_house_historical_2020} with the that, between 1970 and 1980, the centre of mass looked as thought it might turn around and head back toward the US once more. The high level of investments in research had begun to pay off and science was riding high in the public psyche in the US in this period.

Yet, despite the payoff from the space race and the beginning of the computer age, spearheaded by silicon valley in the US, the path of the centre of mass resumed its trajectory toward Europe in the 1980s.  The speed of transit of the centre of mass has remained about the same since 1990s, but this conceals a complex set of forces behind this motion: The rise of Japan as an industrial and research power; the emergence of the professionalisation of research in the UK; the creation of a Europe-wide research strategy embodied in the creation of the European Research Council and centralised strategic funding from the framework program grants and the Horizons 2020 program; and, since 2000, the rise of China as both a major economy and research power.  Indeed, in decades to come we are likely to see the centre of mass travel further as China establishes further and India scales up its research economy.  

\begin{figure}[h]%
\begin{center}
\includegraphics[width=.9\linewidth]{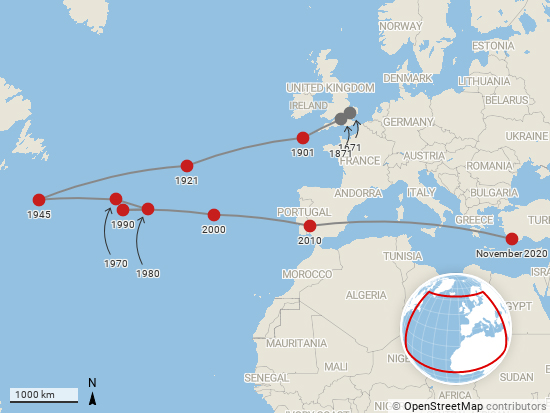}%
\end{center}
\caption{Motion of the centre of mass of research production from 1671 to present day.  The centre of mass calculation is weighted by citations to outputs as described by the Code Listing~\ref{lst:1} and Eqn.~\ref{eq4}.}%
\label{Fig:3}%
\end{figure}

An unweighted calculation shows the clear average centre of production, but it is interesting also to think about different types of weighting.  This should be done with care since the interpretation of such weightings is not trivial. Figure~\ref{Fig:3} shows a similar picture to Figure~\ref{Fig:2}, but this time with each institution's contribution weighted by the fraction of the number of citations associated with the papers written by their affiliated authors. The addition of citation data stablises the path overall, as there is a bias toward the most established research economies.  In this figure, the centre of mass continues to be closest to the US in 1945, but it returns to Europe initially more slowly, and actually turns around, heading back toward the US in the 1980s, before moving once more toward Europe, moving faster than ever, by 2000.  

The speed of movement toward the east has increased significantly over the last 20 years, which is indicative not only of increasing research volumes in China as well as Japan, India, Australia and New Zealand but also the increased citation garnered by those publications.

Additionally, while the range of movement of the centre of mass from east to west is significant, its movement to the south, while being monotonic and more limited in range than the longitudinal motion, is notable by its consistency in the latter half of the 20th Century.  The majority of the world's large cities, and hence most abundant research economies, are in the northern hemisphere.  Yet, the trend is to the South and tracking this motion is sure to be interesting in the future.

\begin{figure}[!h]%
\begin{center}
\includegraphics[width=.9\linewidth]{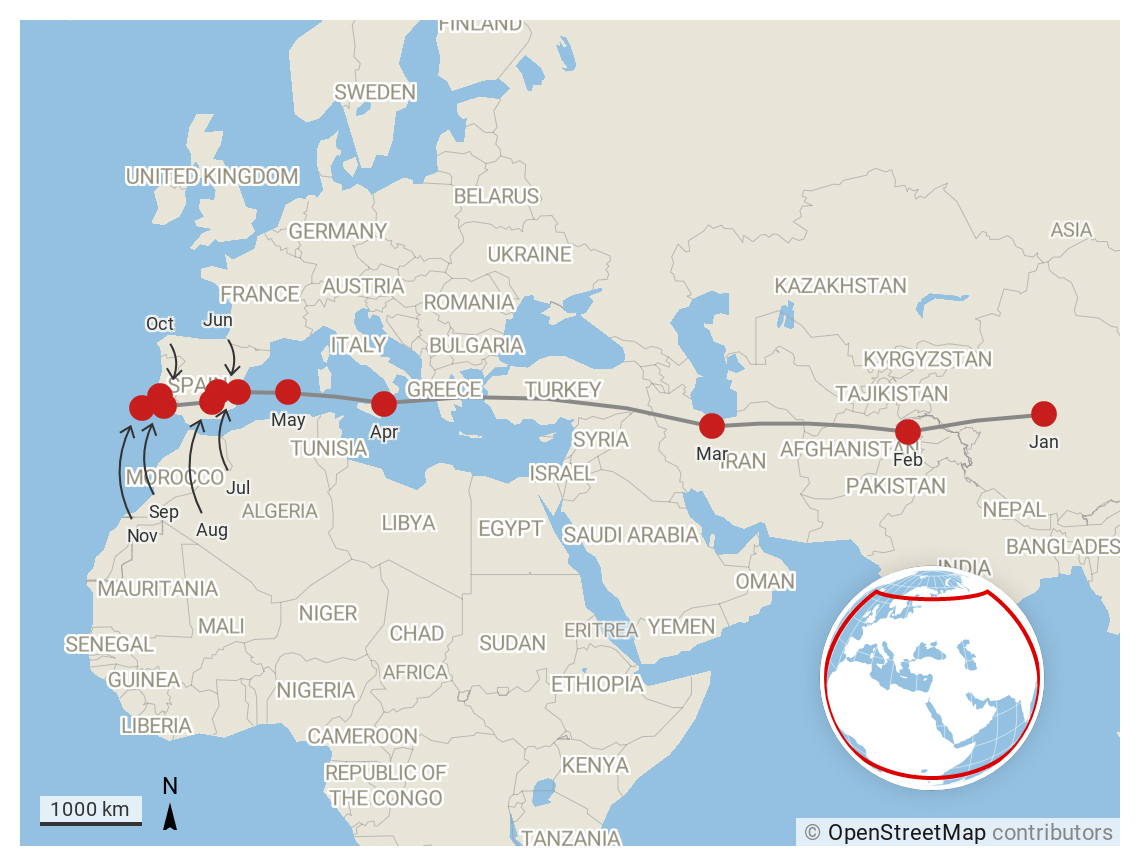}%
\end{center}
\caption{Motion of the centre of mass of research production month by month for COVID-19 publications from January 2020 to December 2020.  The centre of mass calculation is weighted by citations as described by Eqn.~\ref{eq4}.}%
\label{Fig:4}%
\end{figure}

Our third and final narrative is contained in Figure~\ref{Fig:4}, which shows the motion of the citation-weighted centre of mass of COVID-19 research on a monthly basis during 2020.  The number of publications that contribute to each point on the plot is shown in Table~\ref{tab:t1}.  

As news of COVID-19 emerged from Wuhan in China during at the beginning of the year, China's researchers quickly turned their attention to studying the disease.  The fact that the centre of mass of COVID-19 research in January 2020 is located on the Tibetan plateau (paradoxically, quite near to the centre of mass of global economic output in 1CE as calculated in the McKinsey report that originally inspired this line work in this paper) rather than closer to China's research centres is a clear indication that research was already taking place in the international community.  As the year progressed and the virus spread to pandemic migrated West, more and more research organisations in the West turned their attention to COVID-19 research.  The shift in the centre of mass of global research production and the speed at which this happened is easy to see from Fig.~\ref{Fig:4}.

Maps hold a special place in human storytelling and hence are a powerful means by which we can relate to data.  The use of such maps does not come without baggage - such visualisations hide many facets.  However, they are impactful and, we believe that the simplicity of the technology that we've demonstrated in this short article shows great promise as a tool to illustrate trends in academic research.

\section{Discussion}
\label{Sec:Discussion}
\subsection{A new world of analysis}
\begin{figure*}[hbt]%
\begin{center}
\includegraphics[width=\linewidth]{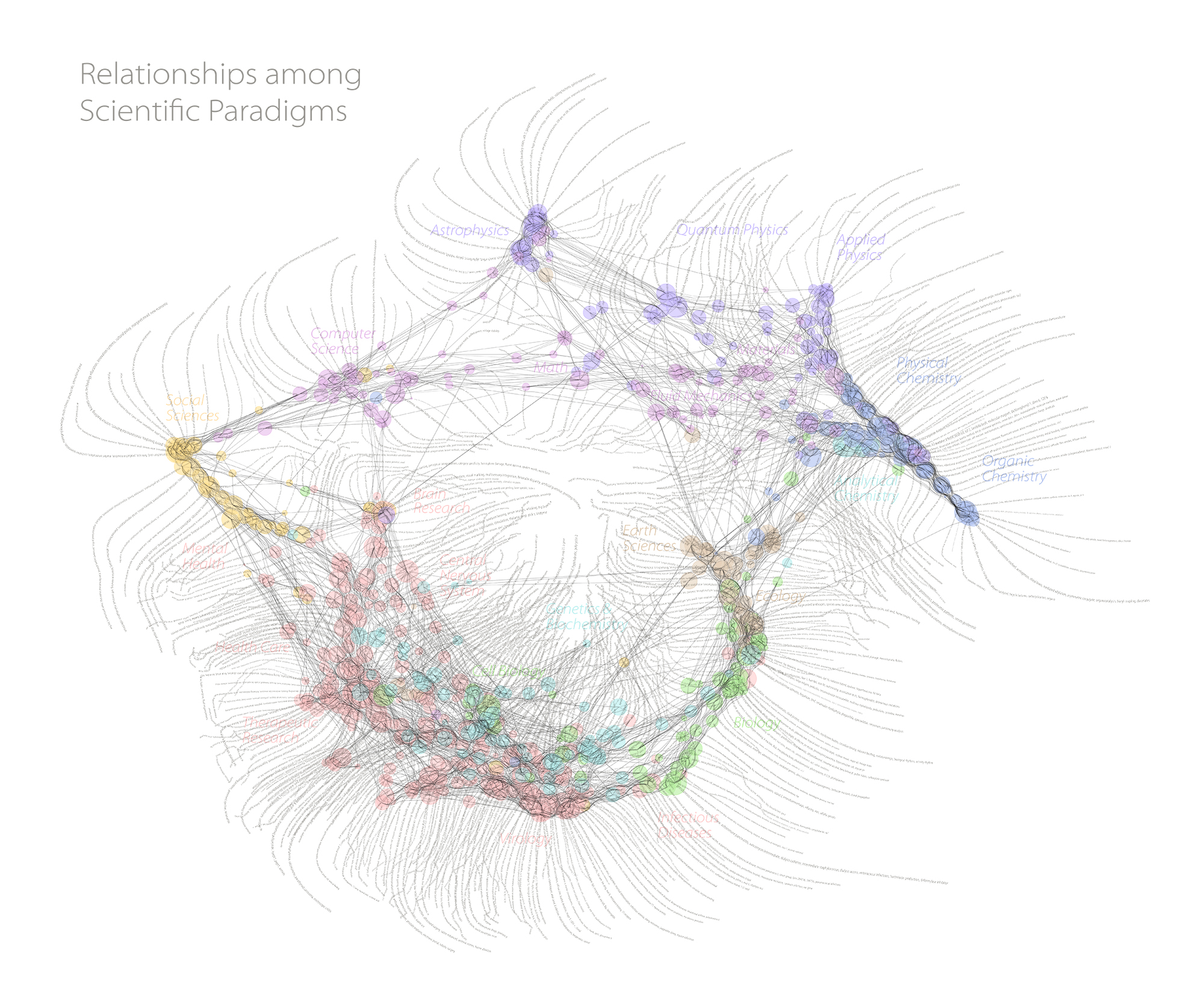}%
\end{center}
\caption{One of the first visualisations of research that made use of a full global dataset.  While ``calculated'', a significant amount of manual work was needed to make this beautiful visualisation, which ensures that detailed data is married with a meaningful visualisation. \emph{Reproduced with kind permission of W B Paley.}}%
\label{Fig:5}%
\end{figure*}

In a recent book \cite{goldin_terra_2020} produce a set of compelling maps with associated narratives.  We have tried to take the same approach in our Results section in order to showcase how these maps may lead to inquiry and contextual interpretation beyond the standard work of analysts.  We have also shown how responsive and immediate these analyses can be - not only adding an interesting thread to historical discourse but allowing us to see emergent trends in real time.  We believe that this type of thinking is well understood by many in the scientometric community, as evidenced by the attention received by the work of W. B. Paley (Fig.~\ref{Fig:5}) and others who originally pioneered research cartography.  One of the enduring challenges of automated data visualisation is the ability to optimise layout and preserve information.  In general, it is not possible to reach the level at which this is done in Paley et al's work.  However, in making it easier to create visualisations on the fly, while we give up the data transparency that Paley aspires to, we are able to add speed of iteration so that a visualisation can be used in an actionable manner.

It is widely recognised that data visualisation is a powerful tool for contextualisation and interpretation \citep{tufte_visual_2001,dick_infographic_2020,rendgen_minard_2018}.  The analysis presented in this paper aims to make three points: Firstly, that data accessibility is a partner to data quality and an important part of how data may be deployed to gain insight; Secondly, that data certain visualisation styles and appraoches have been previously overlooked due to the lack not only of the data accessibility, but also the need for data connectivity through persistent identifiers; Thirdly, that tools like these should not be limited only to the most well funded researchers and that Cloud infrastructure may be an effective mechanism to democratise access to these types of data, tools and interpretation, and hence be a route to superior strategic decision making across the sector.

\subsection{A new world of data}

By introducing the scientific method in his book Novum Organum in 1620, Bacon codified the deep relationship between science and data.  The importance of data is not solely limited to the scientific disciplines, rather data defined by a broad definition has always been part of research, regardless of topic. However, until relatively recently in human history, data has been rare. In the last half century we have seen an explosion in the amount of data made available not only by physical and biological experiments, but also by social experiments and also the emergence of the digital humanities.  We have gone from a poverty of data to an amount of data that cannot be handled by any individual human mind.

As in the wider world of research, scientometrics has seen a rise in data availability over the last twenty to thirty years as the research community has grown and professionalised.  The need for metadata that describes not only the outputs of research but also the process by which they are produced, the broad scholarly record, is now widely acknowledged.  

In the next few years, we are likely to see the amount of metadata collected about a research output increase manyfold, so that the metadata about an object exceeds the data contained within the object.  The ability to scale data systems, share and manipulate data and to summarise it for human consumption in visualisations is becoming critical, as is understanding the biases that are inherent to different visualisation styles.

In moving forward, we argue that critical consideration needs to be given to data accessibility.  Others such as \citep{mons_invest_2020} have argued cogently that investment should be made into research data.  We believe that investment could be helped by introducing a framework such as the one proposed here to support a working definition of data accessibility and good practice. The facets of coverage, structure, nature, context and quality, could form the basis of a helpful rubric for making research data more valuable and accessbile to the community.  There is already a precedent for gaining cross-community collaboration in projects such as I4OC and I4OA as well as structures for use of metrics in DORA and the Leiden Manifesto \citep{hicks_bibliometrics_2015} - is data access another similar area where the community should seek to build principles to ensure the most even playing field?

\section{Future explorations}
The methods explored in this paper can be extended and applied in many different scenarios. It is easy to see how this analysis could be repeated and customised for a variety of geographies (e.g. specific countries or regions), subject areas (e.g. COVID as shown here or Sustainable Development Goals) and timescales. Weighting schemes could include altmetric-based approaches, funding weighting, journal metric-led weighting or any number of different approaches to suit specific needs.  In addition, using Dimensions, parallel analyses could be performed based on grant data, clinical trials data, patent data, pollicy documents or data.  As noted previously, equivalent problems that could make use of similar capabilities and technologies include global heatmapping of specific research activities, the creation of specific custom benchmarks or other metrics to specification and on demand.

We have discussed context as a critical part of research analysis in this paper.  Thus, it is important to highlight the context of the data used in our analyses. Despite the foundational principals behind Dimensions of not editorialising its data holdings, it is still not a universal dataset.  At the current time, not all funding organisations make their data openly available and the publications associated with some geographies and some fields are not held in the DOI registries that have yet been integrated into Dimensions.  As a result, the analysis presented here has flaws and will naturally show an english-language centred view of the world.

In this paper, we have focused on a particular analysis and visualisation style that we have not seen in the scientometric literature before.  We beleive that the lack of use of this style is due to the constraints that we have have outlined.  However, we believe that our underlying argument around data access can be applied also to the production of visualisations such as those offered by VOSviewer, CiteSpace and similar technologies \citep{chen_citespace_2006, colavizza_scientometric_2021}.

We close by commenting that, if adopted broadly, we believe that the Cloud techniques applied in this article can lead to better decision making across academia as analysis can become more iterative and more available across the sector.

\section*{Conflict of Interest Statement}
All authors of this paper are employees of Digital Science, the creator and owner of Dimensions and GRID.

\section*{Author Contributions}

DWH developed the idea for this paper and drafted the manuscript and carried out the visualisation.  SJP developed the implementation of the code, determined the business rules and methodology for the data extraction.  Both co-authors edited and reviewed the manuscript.

\bibliographystyle{frontiersinSCNS_ENG_HUMS} % for Science, Engineering and Humanities and Social Sciences articles, for Humanities and Social Sciences articles please include page numbers in the in-text citations
\bibliography{map}

\end{document}